\documentstyle[aps,epsf,preprint,prbbib,eqsecnum]{revtex}
\begin{document}
\large

\title{Large-scale properties of passive scalar advection}
\author{E. Balkovsky$^a$, G. Falkovich$^a$, V. Lebedev$^{a,b}$ and 
M. Lysiansky$^a$}
\address{$^a$ Physics of Complex Systems,
Weizmann Institute of Science,  Rehovot 76100, Israel \\
$^b$ Landau Institute for Theoretical Physics, RAS, 
Kosygina 2, 117940, Moscow, Russia.}
\date{\today}
\maketitle
\vspace{3cm}
Corresponding author: 

Gregory Falkovich

tel. 972-8-9342830, fax 972-8-9344109, e-mail fnfal@wicc.weizmann.ac.il

PACS numbers 47.10.+g, 47.27.-i, 47.27Qb

\begin{abstract}
We consider statistics of the passive scalar on distances much larger than the
pumping scale. Such statistics is determined by statistics of Lagrangian
contraction that is by probabilities of initially distant fluid particles to
come close. At the Batchelor limit of spatially smooth velocity, 
the breakdown of scale invariance is established for scalar statistics.
\end{abstract}
\pacs{PACS numbers 47.10.+g, 47.27.-i, 47.27Qb}


\section*{Introduction}

If an external pumping supplies the fluctuations of the scalar at some scale
$L$, then the advection by a spatially inhomogeneous velocity field produces
scalar fluctuations at all scales, both larger and smaller than $L$. In an
incompressible velocity field, the flux of the scalar variance flows
downscales, this direct cascade is quite well understood by now
\cite{59Bat,68Kra-a,94SS,95GK,95CFKL}. From a general physical viewpoint, it is
also of interest to understand the properties of turbulence at scales larger
than the pumping scale. If only direct cascade exists, one may expect
equilibrium equipartition at large scales with the effective temperature
determined by small-scale turbulence \cite{FNS,95BFLS}. The peculiarity of our
problem is that we consider scalar fluctuations at the scales that are larger
than the scale of excitation yet smaller than the correlation scale of the
velocity field, which provides for mixing of the scalar. Although we find
simultaneous correlation functions of different orders, it is yet unclear if such a
statistics can be described by any thermodynamics-like variational principle..  

Since we are interested in the behavior of the passive scalar on large
scales, the diffusivity can be neglected, so that the properties of the scalar
statistics are solely due to Lagrangian dynamics. In a turbulent flow, the
distances between fluid particles generally grow with time. The law of such
growth determines the correlation functions of the scalar at the distances
smaller than $L$. For example, the pair correlation function
$\langle\theta({\bbox r})\theta({ 0})\rangle$ is proportional to the average
time two fluid particles spend within the pumping correlation scale. For $r<L$,
that is the time when separation grows from $r$ to $L$. On the contrary, the
scalar statistics at scales larger than $L$ is related to the probabilities of
initially distant particles to come close. Study of the large-scale statistics thus
reveals new information on the properties of Lagrangian dynamics in a random
flow. We shall show below that the statistics of Lagrangian contraction
critically depends on the spatial smoothness of the velocity field. We shall
argue that non-smooth velocity provides for a scale-invariant statistics of a
scalar which is even getting Gaussian at the limiting case of extremely
irregular velocity. Contrary, the statistics is rather peculiar at spatially
smooth random flow (so-called Batchelor limit): it demonstrates strong
intermittency and non-Gaussianity at large scales. Another unexpected feature
of the scalar statistics in this limit is a total breakdown of scale
invariance: not only the scaling exponents are anomalous (i.e. don't grow
linearly with the order of correlation function) yet even any given correlation
function is not generally scale invariant (that is the scaling exponents depend
on the angles between the vectors connecting the points).

The paper is organized as follows. We introduce the problem and discuss the
results that could be understood qualitatively in Sect. \ref{qual}. These
results are supported by straightforward calculations within the framework of
Kraichnan model \cite{68Kra-a}, presented in Sects.  \ref{An}-\ref{thrthr}. 
We briefly describe the case of non-smooth velocity in Sect.
\ref{ns}. We consider arbitrary space dimensionality ${d}$. Two-dimensional
case deserves separate consideration due to an additional degeneracy.

\section{Qualitative Description}
\label{qual}

The evolution of the passive scalar $\theta({\bbox r},t)$ under the action of
velocity ${\bbox v}({\bbox r},t)$ and pumping $\phi({\bbox r},t)$ is described
by the equation
\begin{equation}
  \partial_t\theta+{\bbox v}\nabla\theta=\phi \,.
  \label{ga1}
\end{equation}
Let us introduce Lagrangian trajectories ${\bbox\varrho}({\bbox r},t)$ 
determined by the equation 
$\partial_t{\bbox\varrho}={\bbox v}(t,{\bbox\varrho})$ and by the initial 
condition ${\bbox\varrho}(0,{\bbox r})={\bbox r}$. Next, introducing
$\tilde\theta(t,{\bbox r})=\theta(t,{\bbox\varrho})$ we rewrite 
(\ref{ga1}) as $\partial_t\tilde\theta=\phi$, which gives the formal solution
\begin{equation}
  \theta(0,{\bbox r})=\int_{-\infty}^0\!\! {\rm d}t\, 
    \phi(t,{\bbox\varrho}) \,.
\label{ga3} \end{equation}
Here we have taken into account that at $t=0$ the functions $\theta$ and
$\tilde\theta$ coincide. 

Both ${\bbox v}$ and $\phi$ are assumed to be random functions of time and
space.  We will examine $n$-point correlation functions of the passive scalar
$F_n({\bbox r}_1,\dots,{\bbox r}_n) =\langle\theta({\bbox
r}_1)\dots\theta({\bbox r}_n)\rangle$, averaged over both the statistics of the
advecting velocity ${\bbox v}$ and of the pumping $\phi$. Since our main
interest here is to study the scalar statistics on large distances and time
scales, then without lost of generality we may consider pumping statistics to
be close to white Gaussian
\begin{equation}
  \langle\phi(t_1,{\bbox r}_1)\phi(t_2,{\bbox r}_2)\rangle
    =\delta(t_1-t_2)\chi\left(|{\bbox r}_1-{\bbox r}_2|\right) \,.
  \label{gaa1}
\end{equation}
Here $\chi$ is assumed to decay on a scale $L$. One can treat a deviation
from Gaussianity by introducing the three-point pumping correlation function 
\begin{eqnarray}&&
  \langle\phi(t_1,{\bbox r}_1)\phi(t_2,{\bbox r}_2)
\phi(t_3,{\bbox r}_3)\rangle\label{three} \\ &&=\delta(t_1-t_2)\delta(t_1-t_3)
\chi_3\left(|{\bbox r}_1-{\bbox r}_2|,|{\bbox r}_1-{\bbox r}_3|,
|{\bbox r}_2-{\bbox r}_3|\right) \,,
  \nonumber \end{eqnarray}
where $\chi_3$ is supposed to have the same characteristic
length $L$ as $\chi$. Note, that even when $\chi_3$ introduces a small
correction to the Gaussian statistics of the source, it produces a new
effect, making the odd correlation functions of the scalar non-zero.
The correlation functions can be represented as
\begin{eqnarray} 
F_{2n}=&&\int\limits_{-\infty}^0 {\rm d}t_1 \dots 
\int\limits_{-\infty}^0 {\rm d}t_n
\left\langle \chi[R_{12}(t_1)] \dots 
\chi[R_{2n-1,2n}(t_n)]\right\rangle +\dots \,,
\label{even} \\ 
F_{2n+1}=&&\int\limits_{-\infty}^0 {\rm d}t_1 \dots 
\int\limits_{-\infty}^0 {\rm d}t_n
\biggl\langle\chi_3[R_{12}(t_1),R_{13}(t_1),R_{23}(t_1)]\nonumber\\ && \times
\chi[R_{45}(t_2)] \dots\chi[R_{2n,2n+1}(t_n)]\biggr\rangle +\dots \,,
\label{odd} \end{eqnarray} 
where angular brackets mean averaging over the statistics of the velocity
and one should perform summation over all sets of the pairs of the points
${\bbox r}_i$. Using (\ref{ga3}) we have written the correlation
functions in terms of the Lagrangian separations
\begin{equation}
R_{ij}(t)=|{\bbox\varrho}(t,{\bbox r}_i)
-{\bbox\varrho}(t,{\bbox r}_j)| \,.
\label{sep} \end{equation}
Most of this paper is concerned with the case where the velocity field 
can be considered spatially smooth, which means we can write
\begin{eqnarray} &&
v_\alpha(t,{\bbox r}_1)-v_\alpha(t,{\bbox r}_2)
=\sigma_{\alpha\beta}(t)[r_{1\beta}-r_{2\beta}] \,.
\label{veloc} \end{eqnarray}
Here $\sigma_{\alpha\beta}$ is the random strain matrix depending only on time.
At such a velocity field, the distances $R_{ij}(t)$ grow exponentially, the
stretching rate $\lambda(t)=\ln [R(t)/R(0)]$ has Gaussian statistics with
nonzero mean $\bar\lambda$ and with the dispersion decreasing as $t^{-1/2}$ at
time intervals far exceeding the correlation time of $\hat\sigma$.

Let us briefly remind the properties of the small-scale scalar statistics as
they follow from (\ref{even}--\ref{veloc}). When $r\ll L$, the pair correlation
function is proportional to the mean time when $R(t)<L$ so that
$F_2(r)=\chi(0){\bar\lambda}^{-1}\ln(L/r)$ with logarithmic accuracy 
\cite{59Bat,68Kra-a,94FL}. With the same accuracy, the moments with $n\ll \ln(L/r)$ 
are Gaussian at small scales \cite{94SS,95CFKL}.

The situation is drastically different at $r>L$. Now, non-zero correlation at
two distant points appears only when two fluid particles manage to come there
that were in the past within the pumping correlation length. We thus have to
estimate the probability for the vector ${\bbox R}(t)$ that was once within the
pumping correlation length $L$ to come exactly to the prescribed point ${\bbox
r}$ which is far away. Since the volume is conserved, then all the particles
from the pumping volume $L^d$ will evolve in such a way to be stretched in a
narrow strip with the length $r$.  Assuming ergodicity
[which requires that the stretching time $\bar\lambda^{-1}\ln(r/L)$ is much
larger than the strain correlation time] we thus
come to the conclusion that the probability that two points separated by $r$
belong to a ``piece'' of scalar originated from within $L$ is given by the
volume fraction $(L/r)^d$. That gives the law of the decrease of the two-point
scalar correlation: $F_2\propto r^{-d}$. 

The peculiarity of the smooth velocity field (\ref{veloc}) is that it preserves
straight lines under advection. That makes it easy to determine $r$-dependence
of the correlation function of arbitrary order if all the points lie on a line.
In this case, the history of stretching is the same for all the distances. 
Looking backward in time we may say that when the largest distance between
points was within $L$ then all other distances were as well. Therefore, the
$n$-point correlation function for collinear geometry is determined by the
largest distance: $F_n\propto r^{-d}$. This is true also when different pairs
of points lie on parallel lines. Note that the exponent is $n$-independent
which corresponds to a strong intermittency and an extreme anomalous scaling. 
The fact that for collinear geometry $F_{2n}\gg F_2^n$ is due to strong
correlation of the points along the line.

When we consider an arbitrary geometry, the opposite takes place, namely the
stretching of different non-collinear vectors is generally anti-correlated
because of incompressibility and volume conservation.  Indeed, for a smooth
velocity field there exists a number of invariants, preserved by the flow. A
${\sl d}$-volume $\epsilon_{\alpha_1\alpha_2\ldots\alpha_d}
\rho_1^{\alpha_1}\ldots\rho_d^{\alpha_d}$ is conserved for any ${\sl d}$
Lagrangian trajectories ${\bbox \rho}_i(t)$.  In particular, for ${\sl d}=2$
there are area conservation laws
$\epsilon_{\alpha\beta}\rho_1^\alpha\rho_2^\beta$ for any two vectors relating
three points.  Let us now consider a two-dimensional flow where the
anti-correlation due to area conservation can be easily understood and the
scaling for non-collinear geometry can be readily appreciated.  Since the area
of any triangle is conserved then three points that form a triangle with the
area $s$ much larger than $L^2$ will never come within the pumping correlation
length. Therefore, the triple correlation function 
\begin{eqnarray} &&
F_3(r_{12},r_{13},r_{23})=\int\limits_{-\infty}^0 {\rm d}t
\left\langle\chi_3[R_{12}(t_1),R_{13}(t_1),R_{23}(t_1)]\right\rangle \,
\label{tri1} \end{eqnarray}
is determined by the asymptotic behavior of $\chi_3$ at $r_{ij}\gg L$ that is
very small. For example, if $\chi_3$ decays exponentially then
$F_3\propto\exp(-s/L^2)$. On the other hand, for collinear geometry
$F_3\propto r^{-2}$. We thus see that $F_3$ as a function of the angle
$\vartheta$ between the vectors ${\bbox r}_{12}$ and ${\bbox r}_{13}$ has a
sharp maximum at zero angle and decreases within the interval 
$\vartheta\simeq L^2/r^2\ll1 $.

Similar considerations apply for the fourth-order correlation function
\begin{eqnarray} &&
F_4=\int\limits_{-\infty}^0 {\rm d}t_1\int\limits_{-\infty}^0 {\rm d}t_2
\left\langle \chi[R_{12}(t_1)] 
\chi[R_{34}(t_2)]\right\rangle +\dots \,,
\end{eqnarray} 
where dots stay for all possible permutations of points.  Let us consider the
contribution from the first term. Again, since the area $|{\bbox
R}_{12}\times{\bbox R}_{34}|$ is conserved, the answer is crucially dependent
on the relation between $|{\bbox r}_{34}\times{\bbox r}_{12}|$ and $L^2$. When
$|{\bbox r}_{34}\times{\bbox r}_{12}|\ll L^2$ we have a collinear answer
$F_4\propto r^{-2}$.  Let us now consider the case of non-collinear geometry
and find the probability of an event that during evolution $R_{12}$ became of
the order $L$, and then, at some other moment of time, $R_{34}$ reached $L$
(only such events will contribute into $F_4$).  Note that, unlike the case of
the third-order function, now there is a reducible part in pumping, which makes
$F_4$ nonzero (decaying as power of $r_{ij}$) even when $|{\bbox
r}_{34}\times{\bbox r}_{12}|\gg L^2$.  The probability that $R_{12}$ came to
$L$ is $L^2/r_{12}^2$. Due to area conservation,
there is an anti-correlation between $R_{12}$ and $R_{34}$: if $R_{12}\sim L$,
than $R_{34}\sim r_{12}r_{34}/L$. So probability for $R_{34}$ to come back to
$L$ is $L^2/(r_{12}r_{34}/L)^2 =L^4/r_{12}^2r_{34}^2$.  Therefore, the total
probability can be estimated as $L^6/r^6$, which is much smaller than the naive
Gaussian estimation $L^4/r^4$ while the collinear answer $L^2/r^2$ is much
larger than Gaussian.

That consideration can be readily generalized for arbitrary number of
non-collinear pairs. We expect that $F_{2n}\propto(L/r)^{\Delta_{2n}}$.  In
accordance with (\ref{even}) the separations $R_{ij}$ should be diminished in
the evolution process up to $L$ to produce a non-zero contribution into the
integral.  Suppose that $R_{12}$ is diminished up to $L$. Such process
(explained in the consideration of the pair correlation function) gives the
probability $(L/r_{12})^2$. Next, due to the conservation law of the triangular
areas all other $R_{ij}$ will increase by the factor $r_{12}/L$. Then we should
diminish, say, $R_{34}$ from $r_{34}r_{12}/L$ down to $L$. Such process gives
the probability $L^2/(r_{12}r_{34})$. Due to the conservation law of the
triangular areas other $R_{ij}$ will be larger than their initial values by the
factor $r_{34}/L$ at the moment. Repeating the process we come to the factor
$(L^2/r^2)^{2n-1}$ for the $n$-th order correlation function. Therefore
\begin{eqnarray} &&
\Delta_{2n}=4n-2 \,
\label{lam3} \end{eqnarray}
The above analysis can be generalized for arbitrary geometry. Suppose that
among the separations ${\bbox r}_{ij}$ are parallel vectors (more precisely,
with angles less than $L^2/r^2$). Let us divide ${\bbox r}_{i}$ into sets
consisting of the pairs of points with parallel separations ${\bbox r}_{ij}$.
All points of such set behave as a single separation at the Lagrangian
evolution. Therefore instead of $n$ we should substitute into (\ref{lam3}) the
(minimal) number of the sets.  The estimates, obtained above will be supported
by rigorous calculations in Sec. \ref{DT}.

Unfortunately, not much can be argued qualitatively about the scaling at ${\sl
d}>2$. The crucial point for our considerations in ${\sl d}=2$ was the
conservation of the area. It allowed us to get the correct answers even without
calculations. In other terms, it is related to the fact that there is a single
Lyapunov exponent at two dimensions. When ${\sl d}>2$ we have only the
conservation of the ${\sl d}$-dimensional volumes and hence more freedom in the
dynamics.  Consider, for instance, the three-point correlation function for
non-collinear geometry. Unlike ${\sl d}=2$ we can not assert that it is zero,
since now the area of the triangle is not fixed and can change during the
evolution.  Nevertheless, the anti-correlation between different Lagrangian
trajectories exists, and therefore the answer for the exponent $\Delta_3$
should be larger than $2{\sl d}$ which is the estimate one would get without
the anti-correlation. In the following sections we find that
$\Delta_3=d+(d-1)\sqrt{d/(d-2)}$. This is determined by the hierarchy of
Lyapunov exponents giving the stretching rates at different directions --- 
Sec. \ref{instanton}. Note that in the limit of large $d$ the anti-correlation
should disappear and the answer tends to $2{\sl d}$. The four-point correlation
function is also determined by a joint evolution of two distances and
$\Delta_4=\Delta_3$.

\section{Analytic Calculations}
\label{An}

We do all the calculations assuming the strain to be delta-correlated in time
\begin{equation}
\left\langle\sigma_{\alpha\beta}(t_1)\sigma_{\mu\nu}(t_2)\right\rangle
=D\left[(d+1)\delta_{\alpha\mu}\delta_{\beta\nu}
-\delta_{\alpha\nu}\delta_{\beta\mu}
-\delta_{\alpha\beta}\delta_{\mu\nu}\right]\delta(t_1-t_2) \,.  
\label{ga9} \end{equation} 
The tensorial structure in (\ref{ga9}) is due to isotropy and the
incompressibility condition ${\rm div}\,{\bbox v}={\rm tr}\,\hat\sigma=0$.
Zero correlation time of the strain allows one to derive closed equations
for the correlation functions of the scalar \cite{68Kra-a}:
\begin{eqnarray} &&
D\hat{\cal L}F_{2n}({\bbox r}_{k})
=-\sum\limits_{ij}\chi\left(|{\bbox r}_i-{\bbox r}_j|\right)
F_{2n-2}(t,{\bbox r}_{k'}) \,,
\label{gen1} \\ &&
D\hat{\cal L}F_{2n+1}({\bbox r}_{k})
=-\sum\limits_{ij}\chi\left(|{\bbox r}_i-{\bbox r}_j|\right)
F_{2n-1}(t,{\bbox r}_{k'})
\label{gen2} \\ &&
-\sum\limits_{ijm}\chi_3\left(|{\bbox r}_i-{\bbox r}_j|,
|{\bbox r}_i-{\bbox r}_m|,|{\bbox r}_j-{\bbox r}_m|\right)
F_{2n-2}(t,{\bbox r}_{k''}) \,,
\nonumber \end{eqnarray}
where ${\bbox r}_{k'}$ is the set ${\bbox r}_{k}$ with ${\bbox r}_i$ and
${\bbox r}_j$ excluded and ${\bbox r}_{k''}$ is the set ${\bbox r}_{k}$ with
${\bbox r}_i$, ${\bbox r}_j$ and ${\bbox r}_m$ excluded. The dimensionless
operator $\hat{\cal L}$ is written as follows
\begin{eqnarray} &&
\hat{\cal L}=\sum\limits_{ij}\left[
\frac{d+1}{2}r_{ij}^2\delta^{\alpha\beta}
-r_{ij}^\alpha r_{ij}^\beta \right]
\nabla_i^\alpha\nabla_j^\beta,
\label{gen3} \end{eqnarray}

Eqs. (\ref{gen1},\ref{gen2}) are rather complicated partial differential
equations. We start our analysis from the pair correlation function.

\subsection{Pair correlation function}
\label{sec:pair}

Due to isotropy and translational invariance,  (\ref{gen1}) for the
pair correlation function can be written as
\begin{eqnarray} &&
\frac{(d-1)D}{2}r^{1-d}\partial_r \left(
r^{d+1}\partial_r F_2\right)=-\chi(r) \,,
\label{a9a} \end{eqnarray}
One can easily find a solution of this equation, satisfying the correct
boundary conditions
\begin{eqnarray} &&
F_2(r)=\frac{2}{(d-1)D}\int_r^\infty \frac{{\rm d}x}{x^{d+1}}
\int_0^x{\rm d}y\,\chi(y)y^{d+1}
\nonumber \\ &&
=\frac{2}{d(d-1)D}\left[r^{-d}\int_0^r{\rm d}y\,\chi(y)y^{d-1}
+\int_r^\infty \frac{{\rm d}y}{y}\,\chi(y) \right]\,.
\label{pair} \end{eqnarray}

At $r\gg L$, the function $\chi(r)$ is assumed to decay fast enough
(say, exponentially), and it is possible to neglect the terms related
to the tail of $\chi(r)$ so that
\begin{eqnarray} &&
F_2(r)=\frac{2\bar\chi}{d(d-1)Dr^d}\,.
\label{pasy} \end{eqnarray}
Here $$\bar\chi=\int_0^\infty{\rm d}y\,\chi(y)y^{d-1}$$ is proportional to the
zeroth Fourier harmonics of $\chi(r)$. Estimating $\bar\chi\simeq \chi(0)L^d$
we get $F_2\sim \chi(0)(L/r)^d$. It is important that $\bar\chi$ exists and is
non-zero, otherwise the answer is different.  

\subsection{Collinear Geometry}
\label{Col}

Here we consider a $2n$-th order correlation function of the passive scalar 
regarding that all points ${\bbox r}_1$, \dots, ${\bbox r}_{2n}$
lie on the same line. Then as follows from (\ref{veloc}) during the evolution
${\bbox\varrho}_i$ will remain on a line. The direction of the line can be
characterized by a random unit vector ${\bbox m}(t)$ with the
statistics determined by the equation
\begin{eqnarray} &&
\partial_t m_{\alpha}=\sigma_{\alpha\beta}m_\beta
-m_{\alpha}\zeta \,, \quad
\zeta=\sigma_{\gamma\beta}m_\beta m_\gamma \,,
\label{sta} \end{eqnarray}
following from (\ref{veloc}). For the collinear geometry 
\begin{eqnarray} &&
R_{ij}(\tau)=|{\bbox r}_{i}-{\bbox r}_{j}|
\exp\left\{\int_{\tau}^0 \!\!{\rm d}t\, \zeta(t)\right\} \,.
\label{sta4} \end{eqnarray} 
The statistics of the field $\zeta$ is determined by (\ref{ga9},\ref{sta}) 
which leads to \cite{94CGK,96FKLM}
\begin{eqnarray} &&
\langle\zeta\rangle =\frac{{\sl d}({\sl d}-1)}{2}D \,,
\label{sta1} \\ &&
\left\langle\zeta(t_1)\zeta(t_2)\right\rangle
=D({\sl d}-1)\delta(t_1-t_2) \,.
\label{sta2} \end{eqnarray}
Using the expressions (\ref{sta4},\ref{sta1},\ref{sta2}) we can obtain the 
closed equation for the function $F_{2n}(t,{\bbox r}_k)$:
\begin{eqnarray} &&
\frac{({\sl d}-1)}{2}D \left[{\sl d}
\sum r_{ij}\frac{\partial}{\partial r_{ij}}
+\left(\sum r_{ij}\frac{\partial}{\partial r_{ij}}\right)^2\right]F_{2n}
\nonumber \\ &&
=-\sum\limits_{ij}\chi\left(|{\bbox r}_i-{\bbox r}_j|\right)
F_{2n-2}(t,{\bbox r}_{k'}) \,,
\label{sta3} \end{eqnarray}
where ${\bbox r}_{k'}$ is the set ${\bbox r}_{k}$ with ${\bbox r}_i$ and
${\bbox r}_j$ excluded.

Let us parameterize the points ${\bbox r}_i$ like
\begin{equation}
{\bbox r}_i={\bbox r}_1+ e^\xi{\bbox n}l_i \,,
\label{sta4a} \end{equation}
where ${\bbox n}$ is a unit vector and $l_i$ are some coefficients.
Then, the equation (\ref{sta3}) can be rewritten:
\begin{equation}
\frac{({\sl d}-1)}{2}D 
\left({\sl d}\partial_\xi+\partial_\xi^2\right)F_{2n}
\left(e^\xi{\bbox r}_i\right)
=-\sum\limits_{ij}\chi\left(e^\xi|{\bbox r}_i-{\bbox r}_j|\right)
F_{2n-2}\left(e^\xi{\bbox r}_{k'}\right) \,.
\label{sta5} \end{equation}
This is an ordinary differential equation which has to be solved with the
following boundary conditions: $F_{2n}\left(e^\xi{\bbox r}_i\right)$
tends to zero if $\xi\to+\infty$ and remains finite if $\xi\to-\infty$. 
The solution is
\begin{eqnarray} &&
F_{2n}({\bbox r}_i)=\frac{2}{d(d-1)D}
\int_{-\infty}^{+\infty}{\rm d}\xi\,
\exp\left[-\frac{d}{2}(|\xi|-\xi)\right]
\nonumber \\ &&
\times\sum\limits_{ij}\chi\left(e^\xi|{\bbox r}_i-{\bbox r}_j|\right)
F_{2n-2}\left(e^\xi{\bbox r}_{k'}\right) \,.
\label{sup4} \end{eqnarray}
If the separations $e^\xi|{\bbox r}_i-{\bbox r}_j|$ are much larger than $L$
then the right-hand side of (\ref{sta5}) can be neglected and we conclude that
$F_{2n}\propto\exp(-d\xi)$. Thus we deal with an extremely strong intermittency
when the scaling exponents are independent of $n$. If all separations are of
the same order $r$ then we get from (\ref{sta5}) an estimate
\begin{equation}
F_{2n}\sim\left(\frac{P_2}{D}\right)^n (L/r)^{d} \,.
\label{sta6} \end{equation}
Note that if the distances strongly differ then it follows from (\ref{sup4})
that it is the largest distance that gives the main contribution into
(\ref{sta6}).  

The analogous procedure can be applied to the odd correlation functions of
the passive scalar $\theta$. The only difference is that now we should take
into account also the third-order correlation function of the pumping. Then
we get 
\begin{eqnarray} &&
\frac{({\sl d}-1)}{2}D \left[{\sl d}
\sum r_{ij}\frac{\partial}{\partial r_{ij}}
+\left(\sum r_{ij}\frac{\partial}{\partial r_{ij}}\right)^2
\right]F_{2n+1}
\nonumber \\ &&
=-\sum\limits_{ij}\chi\left(|{\bbox r}_i-{\bbox r}_j|\right)
F_{2n-1}(t,{\bbox r}_{k'})
\label{sta7} \\ &&
-\sum\limits_{ijm}\chi_3\left(|{\bbox r}_i-{\bbox r}_j|,
|{\bbox r}_i-{\bbox r}_m|,|{\bbox r}_j-{\bbox r}_m|\right)
F_{2n-2}(t,{\bbox r}_{k''}) \,,
\nonumber \end{eqnarray}
where ${\bbox r}_{k'}$ is the set ${\bbox r}_{k}$ with ${\bbox r}_i$ and
${\bbox r}_j$ excluded and ${\bbox r}_{k''}$ is the set ${\bbox r}_{k}$ with
${\bbox r}_i$, ${\bbox r}_j$ and ${\bbox r}_m$ excluded. Considering all the
separations of the order of $r$ we get from (\ref{sta7})
\begin{equation}
F_{2n+1}\sim \frac{P_3}{D}
\left(\frac{P_2}{D}\right)^{n-1}
\left(\frac{L}{r}\right)^{d} \,,
\label{sta8} \end{equation}
where $P_3=\chi_3(0,0,0)$. The same $r$-dependence of the odd correlation 
functions as in (\ref{sta6}) is accounted by the the same structure of the
the differential operator in the left-hand sides of (\ref{sta3}) and 
(\ref{sta7}).

\section{Dimensionality Two}
\label{DT}

As we mentioned above, the $2d$ case needs a separate consideration because of
an additional degeneracy of equations for the correlation functions of the
passive scalar. The degeneracy is associated with the area conservation law of
any triangle, vertices of which move along Lagrangian trajectories.

\subsection{Triple correlation function}

As explained in Sec. {\ref{qual}, the three-point correlation function has a
sharp peak for the collinear geometry, whereas for the general position of
points the answer is determined by the tails of the pumping function and is
non-universal. Therefore, only the collinear answer is of interest which has
been already obtained in Sec. \ref{Col}.  Here, we just re-derive the result in
a systematic way, starting directly from the equation $D\hat{\cal L}
F_3=-\chi_3$. Introducing the variables
\begin{eqnarray} &&
x_1=\frac{r_{13}}{r_{12}}\cos\vartheta,\quad 
x_2=\frac{r_{13}}{r_{12}}\sin\vartheta,\quad 
s=r_{12}r_{13}\sin\vartheta \,,
\label{dir}\end{eqnarray}
the operator $\hat{\cal L}$ (\ref{gen3}) can be recast to the following
simple form \cite{95BCKL,97BFL}
\begin{eqnarray}&&
\hat{\cal L}=2x_2^2(\partial_1^2+\partial_2^2) \,.
\end{eqnarray}
Here $\vartheta$ is the angle between ${\bbox r}_{12}$ and ${\bbox
r}_{13}$, and $s$ is the doubled area of the triangle, with vertices in
${\bbox r}_{1}$, ${\bbox r}_{2}$, and ${\bbox r}_{3}$.
Thus, the solution can be easily found
\cite{95BCKL,97BFL} (see also  (\ref{td}))
\begin{eqnarray}&&
F_3=\int_{-\infty}^{+\infty}\!\!{\rm d}x_1'\int_0^{+\infty}
\frac{{\rm d}x_2'}{x_2'^2}\ln\left[
\frac{(x_1-x_1')^2+(x_2+x_2')^2}{(x_1-x_1')^2+(x_2-x_2')^2}\right]
\frac{\chi_3(r_{12}',r_{23}',r_{31}')}{8\pi D}\,.
\label{tc}\end{eqnarray}
One should substitute into  (\ref{tc}) the transformations inverse to (\ref{dir})
\begin{eqnarray} &&
r_{12}=\sqrt{\frac{s}{x_2}},\quad
r_{13}=\sqrt{\frac{s(x_1^2+x_2^2)}{x_2}},\quad
r_{23}=\sqrt{\frac{s([x_1-1]^2+x_2^2)}{x_2}}
\label{inv1} \end{eqnarray}
and the analogous relations between $r_{12}',r_{23}',r_{31}'$ and $s,x_1',x_2'$.

One can easily check, that if $s\gtrsim L^2$, there is no such values of $x_1'$
and $x_2'$, that $r'_{12}$, $r'_{13}$, and  $r'_{23}$ are smaller than $L$.
Therefore, in this case the value of $F_3$ will be determined by non-universal
behavior of the function $\chi_3$ for the values of its arguments larger than
$L$. Consequently, properties of the correlation function $F_3$ are
non-universal, in agreement with the qualitative discussion of Sec. \ref{qual}.

Let us consider $F_3$ at $s\lesssim L^2$. Since we assumed that both $r_{12}$
and $r_{13}$ are much larger than $L$, the condition $s\lesssim L^2$ gives the
inequality $\vartheta<L^2/(r_{12}r_{13})\ll 1$, that is we consider geometry
close to collinear. In this case, the main contribution into the integral
(\ref{tc}) is made by the region of integration, where all $r'$ are smaller
than $L$. In particular, $r'_{12}\lesssim L$, which allows to estimate
$x_2'\gtrsim s/L^2$, which is the same as $x_2'/x_2\gtrsim r_{12}^2/L^2\gg 1$. 
Therefore, we can expand the resolvent in (\ref{tc}) and write
\begin{eqnarray}&&
F_3=\frac{x_2}{2\pi D}\int
\frac{{\rm d}x_2'}{x_2'}\int
\frac{\chi_3(r'_{12},r'_{13},r'_{23})\,{\rm d}x_1'}{(x_1-x_1')^2+x_2'^2}\,.
\label{tc1}\end{eqnarray}
In the limit $s\ll L^2$, it can be further simplified since the main 
contribution to the integral
(\ref{tc1}) is associated with the region $x_2'\ll {\rm min}(1,x_1)$:
\begin{eqnarray}&&
F_3=\frac{1}{D}\int_0^\infty
\chi_3(r_{12}\xi,r_{13}\xi,r_{23}\xi)\, \xi{\rm d}\xi
\approx \frac{P_3 L^2}{2D\,{\rm max}\,
\left(r_{12}^2,r_{13}^2,r_{23}^2\right)} \,.
\label{third} \end{eqnarray}
The expression (\ref{third}) is in accordance with the estimate (\ref{sta8}).
Note that (\ref{third}) has no singularity when any two points coincide as
long as at least one distance is finite.

\subsection{Four-point correlation function}
\label{sec:fotwo}

In this subsection we derive the result for the four-point correlation function
starting directly from (\ref{gen1}). Again, there are two regimes for which
one can find the answer. For the collinear geometry, the consideration is very
similar to one done in the previous section and reproduces the result
(\ref{sup4}). Here we will find the answer for the non-collinear geometry.
Note that its estimate is already known from Sec. \ref{qual}.
The equation (\ref{gen1}) for the four-point correlation function $F_4$ is
$$-D\hat{\cal L} F_4=\chi(r_{12})F_2(r_{34})+\,{\rm permutations}\ .$$
The property of the operator (\ref{gen3}) (characteristic of the large-scale
advecting velocity) is that the solution of this equation is reducible into
pieces, corresponding to each term in it's right-hand side:
\begin{eqnarray} 
F_4=&&\tilde F_4({\bbox r}_{12},{\bbox r}_{34})
+\tilde F_4({\bbox r}_{34},{\bbox r}_{12})
+\tilde F_4({\bbox r}_{13},{\bbox r}_{24})\nonumber\\&&
+\tilde F_4({\bbox r}_{24},{\bbox r}_{13})
+\tilde F_4({\bbox r}_{14},{\bbox r}_{23})
+\tilde F_4({\bbox r}_{23},{\bbox r}_{14}) \,,
\label{six} \end{eqnarray} 
To find $\tilde F_4$ we should solve the equation
\begin{eqnarray}&&
-D\hat{\cal L}\tilde F_4({\bbox r}_{12},{\bbox r}_{34})
=\chi(r_{12})F_2(r_{34})\,.
\label{fc1}\end{eqnarray}
In terms of the variables ($\vartheta$ is the angle between ${\bbox r}_{12}$ 
and ${\bbox r}_{34}$)
\begin{eqnarray}&&
x_1=\frac{r_{34}}{r_{12}}\cos\vartheta,\quad 
x_2=\frac{r_{34}}{r_{12}}\sin\vartheta,\quad 
s=r_{12}r_{34}\sin\vartheta \,,
\label{dir1}\end{eqnarray}
the operator $\hat{\cal L}$ for $d=2$ is \cite{95BCKL,97BFL}
\begin{eqnarray}&&
\hat{\cal L}=2x_2^2(\partial_1^2+\partial_2^2).
\end{eqnarray}
The solution of (\ref{fc1}) can be written as a double integral
\begin{eqnarray} &&
\tilde F_4=\frac{1}{8\pi D}\int_{-\infty}^{+\infty}\!\!{\rm d}x_1'
\int_0^{+\infty}\frac{{\rm d}x_2'}{x_2'^2}
\chi(r'_{12}) F_2(r'_{34})
\ln\left[\frac{(x_1-x_1')^2+(x_2+x_2')^2}
{(x_1-x_1')^2+(x_2-x_2')^2}\right] \,.
\nonumber\\&&
r'_{12}=\sqrt{\frac{s}{x_2'}},\quad
r'_{34}=\sqrt{\frac{s(x_1'^2+x_2'^2)}{x_2'}}\label{vr1}
\end{eqnarray}
We shall calculate the integral in the limit $s\gg  L^2$ when there
are several simplifications. First, since $r'_{12}\lesssim L$, we can write
$x'_2\gtrsim s/L^2$. Hence, $r'_{34}>\sqrt{sx'_2}\gtrsim s/L\gg L$, and we can
use the asymptotic form (\ref{pasy}) of $F_2$.  Second, like for the
three-point correlation function one can show that $x_2'\gg x_2$, and one can
expand the logarithm.  Finally, we can write
\begin{eqnarray}&&
\tilde F_4=\frac{\chi_2x_2}{4\pi^2 D^2s}
\int_{0}^{+\infty}{\rm d}x_2'\,
\chi\left(\sqrt{\frac{s}{x_2'}}\right)
\int_{-\infty}^{+\infty}\frac{{\rm d}x_1'}
{[x_1'^2+x_2'^2][(x_1'-x_1)^2+x_2'^2]}\,.
\label{fc5} \end{eqnarray}
Integral over $x_1'$ can be easily calculated and we get
\begin{eqnarray}&&
\tilde F_4=\frac{\chi_2x_2}{2\pi D^2s}
\int_{0}^{+\infty}
\frac{{\rm d}x_2'}{x_2'}
\frac{1}{x_1^2+4x_2'^2}\chi\left(\sqrt{\frac{s}{x_2'}}\right)
\end{eqnarray}
If $x_1$ is not anomalously large, we can disregard it in the integrand, and
find
\begin{eqnarray}&&
\tilde F_4=\frac{\chi_2Cx_2}{4\pi D^2 s^3}\,,\quad
C=\int_{0}^{+\infty}\!\!
\chi(\xi)\,\xi^3\,{\rm d}\xi\,.
\label{r6}\end{eqnarray}

\section{Dimensionalities larger than two}
\label{thrthr}

Here we treat correlation functions of the passive scalar for $d>2$.  In this
case, the degeneracy inherent to ${ d}=2$ is absent, and the consideration is
the same for all ${ d}$, which is thus considered as a parameter. Of course,
direct physical meaning can be attributed only to ${\sl d}=3$.

We will calculate the three and four-point correlation functions.  Exactly as
it was for ${\sl d}=2$, the operator $\hat{\cal L}$ has the same form for both
correlation functions. Namely, we have to solve the following equations
\begin{eqnarray}&&
-\hat{\cal L}F_3=\chi_3,\quad -\hat{\cal L}\tilde F_4=\chi(r_{12})F_2(r_{34})\,,
\label{F34}\end{eqnarray}
where $\tilde F_4$ is defined by (\ref{six}).
Then, we can introduce the variables (\ref{dir}) for $F_3$ and (\ref{dir1})
for $\tilde F_4$. In these variables the operator $\hat{\cal L}$ has the
following rather simple form
\begin{eqnarray}&&
\hat{\cal L}={\sl d}x_2^2(\partial_1^2+\partial_2^2)+
({\sl d}-2)(\partial_\tau^2+{\sl d} \partial_\tau)\,,\quad
\tau=\ln \left(s/L^2\right)\,.\nonumber
\end{eqnarray}
Therefore, in order to solve Eqs. (\ref{F34}) we have to find the resolvent 
${\cal R}$ of the operator $\hat{\cal L}$ which satisfies the equation
\begin{eqnarray}&&
-\hat{\cal L}{\cal R}=\delta(x_1-x_1')\delta(x_2-x_2')\delta(\tau-\tau')\,.\nonumber
\end{eqnarray}
and the following boundary conditions: First, ${\cal R}$ should go to zero
when $x_1\to\pm\infty$, $x_2\to+\infty$, $x_2\to 0$, and $\tau\to+\infty$.
Then ${\cal R}$ should tend to a constant at $\tau\to-\infty$. 
It is more convenient to work with Hermitian operators, therefore it is useful
to make a substitution
\begin{eqnarray}&&
{\cal R}=\frac{x_2}{{\sl d}x_2'}
\exp\left[-\frac{{\sl d}}{2}(\tau-\tau')\right]R(x_1,x_1',x_2,x_2',\tau,\tau')\nonumber
\end{eqnarray}
Then we obtain an equation
\begin{eqnarray}&&
x_2^2\frac{\partial^2R}{\partial x_1^2}+
x_2\frac{\partial^2 \left(x_2R\right)}{\partial x_2^2}+\frac{d-2}{d}\left(
\frac{\partial^2 R}{\partial \tau^2}-\frac{d^2}{4}R\right)\nonumber\\&&=
-\delta(x_1-x_1')\delta(x_2-x_2')\delta(\tau-\tau')\nonumber
\end{eqnarray}
It is natural to seek the solution in the following form
\begin{eqnarray}&&
R=\frac{1}{\sqrt{x_2}}\int_{-\infty}^{+\infty}
\frac{dk}{2\pi}
\int_{-\infty}^{+\infty}\frac{d\alpha}{2\pi}
u(k,\alpha,x_2,x_2')e^{ik(x_1-x_1')+i\alpha(\tau-\tau')}
\label{fu}\end{eqnarray}
The function $u$ satisfies the following equation
\begin{eqnarray}&&
x_2^2\frac{\partial^2 u}{\partial x_2^2}+x_2\frac{\partial u}{\partial x_2}-
(k^2x_2^2+\nu^2)u=-\sqrt{x_2'}\delta(x_2-x_2')\,,\nonumber
\end{eqnarray}
where
\begin{equation}
\nu=\sqrt{\frac{({\sl d}-1)^2}{4}+\frac{{\sl d}-2}{{\sl d}}\alpha^2}\,.
\label{nu}\end{equation}
The solution of this equation can be readily expressed in terms of the 
Bessel functions of imaginary argument (see e.g. \cite{GR})
\begin{eqnarray}&&
u=\frac{1}{\sqrt{x_2'}}\left\{
\theta(x_2-x_2') K_\nu(|k|x_2)I_\nu(|k|x_2')+
\theta(x_2'-x_2)K_\nu(|k|x_2')I_\nu(|k|x_2)\right\}
\nonumber\end{eqnarray}
Here $\theta(x)$ is the step function, equal to one if $x>0$ and zero otherwise.

Now we should substitute $u$ back into  (\ref{fu}). Integral over $k$ can be
calculated analytically with the help of Eqs. (6.672, 8.820) from \cite{GR}.
Then we get
\begin{eqnarray}&&
R=\frac{1}{2\sqrt{\pi} x_2x_2'}
\int_{-\infty}^{+\infty}\frac{d\alpha}{2\pi}
\frac{\Gamma(\nu+1/2)}{\Gamma(\nu+1)}\left[
\frac{x_2x_2'}{x_2^2+x_2'^2+(x_1-x_1')^2}
\right]^{\nu+1/2}\nonumber\\&&
\times F\left(
\frac{\nu}{2}+\frac{3}{4},\frac{\nu}{2}+\frac{1}{4};\nu+1;
\left[\frac{2x_2x_2'}{x_2^2+x_2'^2+(x_1-x_1')^2}\right]^2
\right)\exp\left[i\alpha(\tau-\tau')\right]
\label{oal}\end{eqnarray}
Here $F(\alpha,\beta;\gamma;x)$ is the hypergeometric function and
$\Gamma(x)$ is the Euler Gamma function.

In ${\sl d}=2$, the integral over $\alpha$ is trivial, and the resolvent can
be easily reproduced \cite{95BCKL,97BFL}
\begin{eqnarray}&&
{\cal R}=\frac{1}{8\pi x_2'^2}\ln\left[
\frac{(x_1-x_1')^2+(x_2-x_2')^2}{(x_1-x_1')^2+(x_2+x_2')^2}
\right]\delta(\tau-\tau')
\label{td}\end{eqnarray}
Convolution of the resolvent with the right-hand sides of Eqs. (\ref{F34})
depends on the properties of the sources. Therefore, below we consider
separately both correlation functions.

\subsection{Three-point correlation function}

The solution of  (\ref{F34}) for the three-point correlation function
can be written in the following form
\begin{eqnarray} 
F_3
=&&\int_0^{\infty} {\rm d}x_2'
\int_{-\infty}^{+\infty}{\rm d}x_1'
\int_{-\infty}^{+\infty}{\rm d}\tau'\,
{\cal R}\left(x_1,x_1',x_2,x_2',\tau,\tau'\right)\,
\frac{\chi_3(r'_{12},r'_{13},r'_{23})}{D}.
\label{up1} \end{eqnarray}
The relations between the variables $r'_{12},r'_{13},r'_{23}$ and
$\tau',\eta',\vartheta'$ are as in (\ref{inv1}). Remind, that $\tau=\ln s/L^2$.

Like it was in ${\sl d}=2$, the behavior of $F_3$ is very different for the
cases of $\vartheta\ll 1$ and $\vartheta\sim 1$. Let us first consider the case
of not very small angles, namely $\vartheta\gg L^2/(r_{12}r_{13})$. Since both
$r_{12}$ and $r_{13}$ are much larger than $L$, the area $s$ of the triangle is
much larger than $L^2$, which means that $\tau\gg 1$. On the other hand, since
$\chi_3$ decreases very rapidly when any of $r_{ij}$ is larger than $L$, the
area $s'$ can not be much larger than $L^2$. Therefore, $\tau'$ is of the order
unity in the integral (\ref{up1}). Thus, we see that $\tau-\tau'$ is always
positive and much larger than unity. On the other hand, from the condition
$\vartheta\gg L^2/r^2$, it is easy to check that for a typical configuration,
contributing into (\ref{oal}) the condition $\ln A\ll \tau$ holds. 
Therefore, shifting the contour of integration in (\ref{oal}) into the upper
half-plane, we will meet a branch point of the integrand, that originates from
$\nu$ and is situated at $\alpha=i({\sl d}-1)\sqrt{d/(d-2)}$. Because of the
large value of $\tau$, the integral will be determined by a near vicinity of
the branch point. Therefore, we can write
\begin{eqnarray}&&
{\cal R}\propto\exp\left[-\frac{\Delta_3}{2}\tau\right]
\label{pp21}\\ &&
\Delta_3=d+(d-1)\sqrt{d/(d-2)} \,.
\label{up8} 
\end{eqnarray}
Substituting the expression into (\ref{up1}) we get
\begin{eqnarray} &&
F_3\sim ({P_3}/{D})\left(L/r\right)^{\Delta_3} \,, 
\label{up9} \end{eqnarray}
Note that $\Delta_3>2d$. 

Let us now consider the limit $\vartheta\ll L^2/r^2$ and reproduce the
collinear result. In this case, as we shall see, the main contribution to  
(\ref{up1}) is made by $\vartheta'\ll1$. The resolvent ${\cal R}$ in the limit
$\vartheta,\vartheta'\ll 1$ can be found  directly from the representation
(\ref{fu}). The smallness of the angles implies that $x_2,x_2'\ll1$.
Using the asymptotic expansion of the Bessel functions one gets
\begin{eqnarray}&&
u=\frac{1}{2\nu\sqrt{x_2'}}\left\{
\theta(x_2-x_2')\left(\frac{x_2'}{x_2}\right)^\nu+
\theta(x_2'-x_2)\left(\frac{x_2}{x_2'}\right)^\nu\right\}
\end{eqnarray}
We see, that in the main approximation the $k$-dependence disappears of $u$
and we can integrate over $k$ in (\ref{fu}). Then we get
\begin{eqnarray}&&
R=\frac{\delta(x_1-x_1')}{2\sqrt{{x_2}{x_2'}}}\int_{-\infty}^{\infty}
\frac{d\alpha}{2\pi\nu(\alpha)}
\exp\left[
i\alpha(\tau-\tau')-\nu(\alpha)\left|
\ln\frac{x_2}{x_2'}
\right|\right]
\label{rnu}\end{eqnarray}
The integral (\ref{rnu}) can be calculated analytically
\begin{eqnarray}&&
R=\frac{1}{\pi}\sqrt{\frac{x_2}{x^{'3}_2{\sl d}({\sl d}-2)}}
K_0\left({\cal X}\right)\exp\left[
-\frac{{\sl d}}{2}(\tau-\tau')
\right]\delta(x_1-x_1')
\label{K0}\\&&
{\cal X}=\frac{{\sl d}-1}{2}\sqrt{\frac{{\sl d}}{{\sl d}-2}(\tau-\tau')^2+
\ln^2\frac{x_2}{x'_2}}\,.
\end{eqnarray}
The $\delta$-function forces the ratio $r'_{12}/r'_{13}$ to be
equal to $r_{12}/r_{13}$. Integration over one of the distances, say $r'_{12}$
makes both of $r'$ to be of the order $L$ (we believe that $r_{12}\sim r_{13}$).
Let us consider the integral over the angle $\vartheta'$.
It is easy to see, that the argument of $K_0$ in (\ref{K0}) is always large. 
Therefore we can use the asymptotic form of this function and write
\begin{eqnarray} 
F_3\sim&&\int d\vartheta'\frac{\vartheta^{1/2}}{\vartheta'^{3/2}}\exp\Biggl[-\frac{d}{2}
\ln\left(\frac{r^2}{L^2}\frac{\vartheta}{\vartheta'}\right)
-\frac{d-1}{2}\sqrt{\ln^2\frac{\vartheta}{\vartheta'}+\frac{d}{d-2}\ln^2
\left(\frac{r^2}{L^2}\frac{\vartheta}{\vartheta'}\right)}\Biggr]
\nonumber\end{eqnarray}
The main contribution to the integral is made by the vicinity
of $\vartheta'=\vartheta r^2/L^2$. Hence
$F_3\sim \left(\frac{L}{r}\right)^d$.
From the assumption $\vartheta'\ll 1$, we see that there should be
$\vartheta\ll L^2/r^2$, otherwise the main contribution comes from
$\vartheta'\sim 1$, and the expression for the resolvent (\ref{K0}) is
inapplicable.

\subsection{Four-point Correlation Function}

From  (\ref{F34}) it follows that the answer for the four-point correlation
function can be written in the form (\ref{six}) where
\begin{eqnarray} 
\tilde F_4({\bbox r}_{12},{\bbox r}_{34})
=&&\frac{1}{D}\int_0^{\infty} {\rm d}x_2'
\int_{-\infty}^{+\infty}{\rm d}x_1'
\int_{-\infty}^{+\infty}{\rm d}\tau'\nonumber\\&&\times
{\cal R}(x_1,x'_1,x_2,x'_2,\tau,\tau')\,\chi(r'_{12})F_2(r'_{34}) \,,
\label{ip1}\end{eqnarray}
The variables $r'_{12}$ and $r'_{34}$ are expressed via $x'_1$, $x'_2$, and
$\tau'$ by  (\ref{vr1}). The analogous relations hold for the variables
$r_{12}$ and $r_{34}$. It can be more convenient for the present purposes to
pass from the integration over $x'_1$, $x'_2$, and $\tau'$ to the integration
over $\vartheta'$, $r'_{12}$, and $r'_{34}$. Then (\ref{ip1}) can be replaced by
\begin{eqnarray}&&
\tilde F_4=\frac{1}{2{\sl d} \pi^{3/2}}\int_{-\infty}^\infty 
\frac{dr'_{12}}{r'_{12}}
\chi\left(r'_{12}\right)
\int_{-\infty}^\infty \frac{dr'_{34}}{r'_{34}}  F_2\left(r'_{34}\right)
\int_0^\pi \frac{d\vartheta'}{\sin^2\vartheta'}
\int_{-\infty}^{\infty}d\alpha\nonumber\\&&\times\frac{\Gamma(\nu+1/2)}{\Gamma(\nu+1)}
\left(\frac{A}{2}\right)^{\nu+1/2}
F\left(\frac{\nu}{2}+\frac{3}{4},\frac{\nu}{2}+\frac{1}{4};\nu+1;A^2\right)
\left(
\frac{r_{12}r_{34}\sin\vartheta}{r'_{12}r'_{34}\sin\vartheta'}
\right)^{i\alpha-{\sl d}/2}
\nonumber\end{eqnarray}
The expression for $A$ can be written as follows
\begin{eqnarray}&&
A=\frac{2\sin\vartheta\sin\vartheta'}{r_{12}r'_{34}/(r'_{12}r_{34})+
r'_{12}r_{34}/(r_{12}r'_{34})-2\cos\vartheta\cos\vartheta'}\,.
\end{eqnarray}

The case $\vartheta\ll L^2/r^2$ can be analyzed in a way, presented in the
previous subsection for the three-point correlation function, leading to the
expression (\ref{sup4}) and to the law (\ref{sta6}). Below we analyze the case
$\vartheta\gg L^2/r^2$.  In this case, there are two different regions in the
integral over $\xi'_2$ in the integral making contribution to $\tilde F_4$. The
first region is determined by the inequality $\xi'_2\lesssim 1$, which
corresponds to $r'_{34}\simeq L$.  By the order of magnitude, this contribution
is equal to the three-point correlation function, because in this case, the
value of $\tau'$ in (\ref{ip1}) is of the order of unity, while $\tau\gg 1$ and
therefore all the arguments presented in the previous subsection are valid. 
The second region is defined by the condition $\xi'_2\gtrsim 1$. It
corresponds to large values of $r'_{34}$, for which we can use the asymptotic
behavior of $F_2$, given by (\ref{pasy}).  Another simplification is that in
this region $r'_{34}\gg r'_{12}$ and therefore we can believe
\begin{eqnarray}&&
A\approx 2\sin\vartheta\sin\vartheta' \,\,
\frac{r_{34}r'_{12}}{r_{12}r'_{34}}\ll 1
\end{eqnarray}
Then, we can put the hypergeometric function to be equal to $1$, and write
\begin{eqnarray}&&
\tilde F_4\approx\frac{2\chi_d}{\pi^{3/2}{\sl d}^2({\sl d}-1)D^2}
\int d\alpha\frac{\Gamma(\nu+1/2)}{\Gamma(\nu+1)}
\left(\frac{r_{34}}{r_{12}}\sin\vartheta\right)^{\nu+1/2}
\left(r_{12}r_{34}\sin\vartheta\right)^{i\alpha-d/2}\nonumber\\&&\times
\int\frac{du}{u}\chi(u)\,u^{\nu+1/2-i\alpha+d/2}
\int\frac{dv}{v}v^{-d/2-\nu-1/2-i\alpha}
\int d\vartheta'\left(\sin\vartheta'\right)^{\nu-3/2-i\alpha+d/2}\,.\nonumber
\end{eqnarray}
The integral over angle $\vartheta'$ can be easily expressed via
the Euler $\Gamma$-functions. The
integral over $v$ should be calculated with the cut-off on $L$.
Therefore, we get the following integral over $\alpha$
\begin{eqnarray}&&
\tilde F_4\propto
\int d\alpha\frac{\Gamma(\nu+1/2)}{\Gamma(\nu+1)}
\frac{\Gamma(\nu/2-1/4-i\alpha/2+{\sl d}/4)}
{\Gamma(\nu/2+1/4-i\alpha/2+{\sl d}/4)}
\left(\frac{r_{12}r_{34}}{L^2}\sin\vartheta\right)^{i\alpha-d/2}\nonumber\\&&\times
{(r_{34}\sin\vartheta/r_{12})^{\nu+1/2}\over{{\sl d}/2+\nu+1/2+i\alpha}}
\int\frac{du}{u}\chi(u)\,
\left(\frac{u}{L^2}\right)^{\nu+1/2-i\alpha+d/2}
\label{pol}\end{eqnarray}
Again, we can shift the contour of integration up to the branch point
determined by the same condition $\nu=0$, and we get the same answer
as for the three-point correlation function. 
Thus, both contributions possess an identical $r$-dependence giving
\begin{eqnarray} &&
F_4\sim\frac{P_2^2}{D^2}\left(\frac{L}{r}\right)^{\Delta_3} \,,
\label{four} \end{eqnarray} 
with the same exponent (\ref{up8}). 

The contribution (\ref{four}) to $F_4$ is the leading one only if $d>\sqrt 2+1$.
If $d<\sqrt 2+1$, then along with (\ref{four}) there appears an additional
contribution into $F_4$ due to a pole of the integrand in  (\ref{pol}).
It originates from the zero of the denominator ${\sl d}/2+\nu+1/2+i\alpha$,
existing only at $d<\sqrt 2+1$. The contribution behaves like
\begin{eqnarray} &&
\propto\left(\frac{L}{r}\right)^{d(d+1)/(d-1)} \,.
\label{ip11} \end{eqnarray}
Comparing the law (\ref{four}) with (\ref{ip11}), we conclude that
in the region of its existence, that is at ${\sl d}<1+\sqrt{2}$, the term
(\ref{ip11}) is the leading contribution to
$\tilde F_{4}({\bbox r}_{12},{\bbox r}_{34})$. Particularly, this is the case
in $d=2$, where the contribution (\ref{ip11}) behaves as $(L/r)^6$,
in accordance with  (\ref{r6}).

\subsection{Instanton}
\label{instanton}

To understand better the underlying Lagrangian dynamics let us outline briefly
another way of calculation, based on the fact that they are rare events that
contributes the correlation functions at large scales.  Therefore, some kind of
instanton formalism can be applied, the main task here is to recognize the
relevant degrees of freedom.  In this way, we shall establish a relation
between the scaling exponent $\Delta_3$ and the Lyapunov exponents of the
smooth flow.  The Lagrangian distances ${\bbox R}_{ij}$ are all determined by a
single matrix $W=T\exp(\int\sigma dt)$ via ${\bbox R}_{ij}=W{\bbox r}_{ij}$. 
To find the correlation functions, we should be able to average over the
statistics of the matrix $W$. The way to do it was proposed in \cite{93Kol}, we
shall follow \cite{98KLS} and write
\begin{eqnarray}&&
W=RT\,,
\end{eqnarray}
where $R$ is an orthogonal matrix and $T$ is an upper-triangular matrix:
$T_{ij}=0$ if $i>j$.
The matrix $R$ can be excluded from the consideration due to isotropy,
then, representing $T$ as
\begin{eqnarray}&&
T_{ii}=\exp(\rho_i),\quad T_{ij}=\exp(\rho_i)\eta_{ij}\quad\mbox{if}\quad i<j
\end{eqnarray}
we can write the action describing the stochastic dynamics of $\rho$
and $\eta$
\begin{eqnarray}&&
{\cal L}=\sum_{i=1}^d m_i\left[
\partial_t\rho_i+D\frac{d(d-2i+1)}{2}\right]+
\frac{\imath D}{2}\left[
d\sum_i m_i^2-\left(\sum_a m_a\right)^2
\right]\nonumber\\&&
+\imath Dd\sum_{i<j}\exp(2\rho_i-2\rho_j)\mu_{ij}^2+
2\imath Dd\sum_{i<k<j}\mu_{ij}\mu_{ik}\exp(2\rho_k-2\rho_i)\eta_{kj}
\nonumber\\&&
+\sum_{i<j}\mu_{ij}\partial_t\eta_{ij}
+\imath Dd\sum_{i<k<m,n}\mu_{im}\mu_{in}
\eta_{km}\eta_{kn}\exp(2\rho_k-2\rho_i)
\label{act}\end{eqnarray}
Here $m_i$ and $\mu_{ij}$ are auxiliary fields, conjugated to $\rho_i$ and
$\eta_{ij}$ respectively.

The variables $\rho_i$ describe stretching of volume elements in the flow,
while $\eta_{ij}$ describe the direction fluctuations of a given vector with
respect to the main axis of the strain matrix $\hat\sigma$.  Note that the
constants $\lambda_i=d(d-2i+1)/2$ entering the action (\ref{act}) are the
Lyapunov exponents.  Using (\ref{act}), we can rewrite $F_3$ (\ref{tri1}) as
follows
\begin{eqnarray}&&
F_3=\int {\cal D}\rho{\cal D}\eta{\cal D}\mu{\cal D}m
\exp\left[i\int_{-\infty}^0 \!\!dt \,{\cal L}+\ln\left(
\int_{-\infty}^0 \!\!dt\,\chi_3\right)\right]
\end{eqnarray}

The variables $\eta_{ij}$ are irrelevant for the
evaluation of the scaling and are only responsible for the angular behavior. 
Since all three $R_{ij}$ always lie on a plane,
we can leave only two of $\rho_i$ and write an effective Lagrangian
\begin{eqnarray}&&
\tilde{\cal L}=m_a(\partial_t\rho_a+\lambda_a)+m_b(\partial_t\rho_b+\lambda_b)
+\frac{\imath D}{2}\left[d(m_a^2+m_b^2)-(m_a+m_b)^2\right]\nonumber
\end{eqnarray}
Then, the dependence on $L/r$ can be extracted from the following expression
\begin{eqnarray}&&
F_3\sim \int {\cal D}\rho_{a,b}{\cal D} m_{a,b}\exp\left[
\imath \int_{-\infty}^0 \!\!dt \,\tilde{\cal L}+\ln\left(
\int_{-\infty}^0 \!\!dt\,\chi_3\right)
\right]\label{f3}
\end{eqnarray}
Now we can write the instanton equations for the extremum of the exponent 
in (\ref{f3})
\begin{eqnarray}&&
\partial_t\rho_a+\lambda_a=-\imath D[(d-1)m_a-m_b]\,,\quad
\partial_t\rho_b+\lambda_b=-\imath D[(d-1)m_b-m_a]\,,\nonumber\\&&
\imath \partial_tm_a=\frac{1}{F}\frac{\partial\chi_3}{\partial\rho_a}\,,\quad
\imath \partial_tm_b=\frac{1}{F}\frac{\partial\chi_3}{\partial\rho_b}\,,
\quad
F=\int_{-\infty}^0 dt\,\chi_3\label{mab}
\end{eqnarray}
The boundary conditions are $m_{a,b}\to 0$ as $t\to -\infty$ and
$\rho_{a,b}=0$ at $t=0$.
Note that the "energy"
\begin{eqnarray}&&
E=\imath (m_a\lambda_a+m_b\lambda_b)-
\frac{D}{2}\left[d(m_a^2+m_b^2)-(m_a+m_b)^2\right]+\frac{1}{F_3}\chi_3
\end{eqnarray}
is a constant. From the boundary conditions we deduce that it should be zero.

Let us explain the qualitative behavior of the solution. We consider the
evolution backwards in time. At small times all $R_{ij}$ are large, and
therefore the derivatives of $\chi_3$ in the right-hand sides of Eqs. 
(\ref{mab}) are zero. Hence $m_{a,b}$ are constants such that $\rho_{a,b}$
diminish and $R_{ij}$ also become smaller. Then, at some moment all $R_{ij}$
become of the order of $L$. Then, derivatives of $\chi_3$ can not be
disregarded and during some short interval of time when $R_{ij}\sim L$ the
values of $m_{a,b}$ will change to vacuum zero values. The derivatives of
$\rho_{a,b}$ change sign, and $R_{ij}$ start to grow. Note, that if only one of
$R_{ij}$ is of the order $L$ and the other are still much larger than $L$, then
$\chi_3$ is small, the derivatives in Eqs. (\ref{mab}) are ineffective, and the
solution will never reach its vacuum stage.  Thus, we should tune the
conditions so that all $R_{ij}$ will become of the order $L$ simultaneously.
Finally, we come to the set of conditions for the initial stage
\begin{eqnarray}&&
\partial_t\rho_a=\lambda=-\lambda_a-\imath D[(d-1)m_a-m_b]\,,\nonumber\\&&
\partial_t\rho_b=\lambda=-\lambda_b-\imath D[(d-1)m_b-m_a]\,,\nonumber\\&&
E=\imath (m_a\lambda_a+m_b\lambda_b)-\frac{D}{2}\left[
d(m_a^2+m_b^2)-(m_a+m_b)^2\right]=0\nonumber
\end{eqnarray}
From here we find the value of $\lambda$
\begin{eqnarray}&&
\lambda=-\frac{1}{2}\sqrt{(\lambda_a+\lambda_b)^2+
\frac{d-2}{d}(\lambda_a-\lambda_b)^2}
\end{eqnarray}
Calculating the action, we find that $F_3\sim (L/r)^{\Delta_3^{a,b}}$ with
\begin{eqnarray}&&
\Delta_3^{a,b}=\frac{\lambda_a+\lambda_b+
\sqrt{(\lambda_a+\lambda_b)^2+\frac{d-1}{d}(\lambda_a-\lambda_b)^2}}{d-2}
\label{dt3}\end{eqnarray}
The value of $\Delta_3^{a,b}$ is minimal (that is $F_3$ is maximal) if we take
the two largest Lyapunov exponents: $\lambda_a=\lambda_1=d(d-1)/2$ and
$\lambda_b=\lambda_2=d(d-3)/2$.  Substituting it into (\ref{dt3}) we reproduce
(\ref{up8}). The instanton found has a long life-time, proportional to
$\ln(r/L)$, therefore the above consideration has to be valid also for a
velocity finite-correlated in time.

\subsection{Discussion}

We thus see that many-point correlation functions are not scale invariant
because of strong angular dependence. One may consider averaging over different
geometries, for instance, integrating over the angle between any two vectors
${\bf r}_{ij}$ keeping $R^2=\sum r_{ij}^2$ in 2d. 
As a result of such averaging, the object
appears which depends on $R$ only and is thus scale invariant. Does such
averaging restores also the normal scaling? One may notice that the main
contribution into the angular integral gives the region of small angles near
collinear peak $\vartheta\lesssim (L/r)^2$; since there are $n-1$ angles in the
$n$-point correlation function then one gets $F_n\propto r^{-2n}$ that is
normal scaling is restored indeed. That means that the increase at small angles
and decrease at large ones (relative to a normal scaling) are of the same
order and both caused by the same mechanism. It is unclear what is the way -
if any - of natural average over geometries that restores normal scaling in
$d>2$. 

We thus discovered an intermittency build-up in the direction opposite to the
cascade. It is instructive to compare this with an intermittency discovered 
at small scales when the cascade flows upscales in a compressible flow {\cite{97CKVa}}.

What will be for a finite correlation of $\hat\sigma$ in time? It is clear
from Sects. \ref{qual},\ref{An} that
the dependence $r^{-d}$ both for the pair correlation function and for the 
correlation function of any order for collinear geometry is valid 
as long as the correlation time of $\hat\sigma$ is much less than
$D^{-1}\ln{(r/L)}$. Under the same assumption, 
all the results obtained in two dimensions will be valid
(up to a single numerical factor in front of any correlation function) for a 
finite-correlated strain as well. As far as higher dimensions are concerned,
it is clear that some anti-correlation between contraction of different
distances will be present, it is likely that it will be governed by the same
exponent $\Delta_3$. Indeed, (\ref{dt3}) has to hold in a finite-correlated
case as well, $\Delta_3$ is thus determined by two largest Lyapunov exponents
which are likely to be proportional to $(d-1)$ and $(d-3)$ respectively.

\section{Non-smooth velocity field}
\label{ns}

The advection of the passive scalar by the non-smooth velocity in the 
framework of the Kraichnan model is described by Gaussian velocity field
with the pair correlation function
$$\langle  u^\alpha(t,{\bf r}) u^\beta(0,0)\rangle=
\delta(t)\left\{V_0
\delta^{\alpha\beta}-r^{-\gamma}
\bigl[(d+1-\gamma)r^2\delta^{\alpha\beta}-(2-\gamma)r^\alpha
r^\beta\bigr]\right\}\ . $$
Here $\gamma$ is a measure of velocity non-smoothness, $0\leq\gamma\leq2$. The
generalization of the equations (\ref{gen1},\ref{gen2}) for nonzero $\gamma$
needs replacing $\hat{\cal L}$ by 
$$\sum_{i,j} r_{ij}^{-\gamma}\bigl[(d+1-\gamma)r_{ij}^2\delta^{\alpha\beta}-
(2-\gamma)r_{ij}^\alpha  r_{ij}^\beta\bigr]
\nabla^\alpha_{i}\nabla^\beta_{j}\ .$$
The pair correlation function is now $F_2(r)\propto r^{\gamma-d}$ 
\cite{95CFKL}. 

An interpretation of the extra factor $r^{\gamma}$ comparing to (\ref{pasy})
is related to the fact that  every Lagrangian distance $R$ generally grow by
a power law $t^{1/\gamma}$ as distinct from an exponential law at $\gamma=0$.
In other words, stretching is uniform in  the logarithm of scale at $\gamma=0$
and decelerating at $\gamma\not=0$.

We cannot yet find the high-order correlation functions at arbitrary $\gamma$.
Fortunately, at the limit of extremely irregular velocity $\gamma=2$ the
operator turns into
$$\hat{\cal L}_0=-{(d-1)}
\left[\left(\sum\nabla_i\right)^2-\sum\nabla_i^2\right] $$
For translationally invariant functions the term
$(\sum\nabla_i)^2$ can be discarded. We are thus left with diffusion equation,
it is straightforward to show that if the pumping is Gaussian then the scalar
statistics is Gaussian too (both at the scales larger and smaller than the
pumping scale). If the pumping is  non-Gaussian, then the scalar statistics
is getting Gaussian at the scales distant  form the pumping scale (odd
moments and cumulants of even moments decrease with the growth of $r/L$
faster than the respective Gaussian moments). The third moment, for instance, 
decreases with $r/L$ faster than $F_2^{3/2}$, this can be shown by
substituting the resolvent ($d>2$)
$${\cal R}=\hat{\cal L}_0^{-1}=-\frac{\Gamma\left(3d/2-1\right)}{2\pi^{3d/2}
D(d-1)}\left[({\bbox x}_1-{\bbox y}_1)^2+
({\bbox x}_2-{\bbox y}_2)^2+({\bbox x}_3-{\bbox y}_3)^2\right]^{1-3d/2}$$
into the expression
\begin{eqnarray}&&
F_3^{(0)}=\int d{\bbox x}_1d{\bbox x}_2d{\bbox x}_3\,
{\cal R}({\bbox x}_1,{\bbox x}_2,{\bbox x}_3,{\bbox y}_1,{\bbox y}_2,
{\bbox y}_3)\,\chi_3(x_{12},x_{13},x_{23})\nonumber\\&&
\approx \frac{C_3}{[y_{12}^2+y_{13}^2+y_{23}^2]^{d-1}}\nonumber\,,\\&&
C_3=\frac{6^{d-1}\Gamma(d-1)}{\pi^d(d-1)}
\int d{\bbox \zeta}d{\bbox \eta}\,
\chi_3\left(2|{\bbox \eta}|, |\sqrt{3}{\bbox \zeta}+{\bbox \eta}|,
|\sqrt{3}{\bbox \zeta}+{\bbox \eta}|\right)
\end{eqnarray}
One sees that $F_3^{(0)}\propto r^{2-2d}$ which is decaying with $r$
much faster than $F_2^{3/2}\propto r^{3-3d/2}$.

At $\gamma<2$ one can directly check that the scalar statistics is non-Gaussian
even for a Gaussian pumping.  
Employing perturbation theory with respect to $\xi=2-\gamma$, 
one may try to prove at least that the scaling is
normal and the angular anomaly is absent at $0<\gamma\leq2$. Let us do this 
for the triple correlation function. The operator $\hat{\cal L}$ to the 
first order in $\xi$ is $\hat{\cal L}=\hat{\cal L}_0 +\xi \hat{\cal L}_1$ with
\begin{eqnarray}&&
\hat{\cal L}_1 =-\sum_{i<j}\left(\delta^{\alpha \beta}
\left[(d-1)\ln r_{ij}+1\right]-
\frac{r_{ij}^{\alpha}r_{ij}^{\beta}}{r_{ij}^{2}}\right)
\nabla_i^\alpha\nabla_j^\beta
\nonumber\end{eqnarray}
Then, we should find $\hat{\cal L}_1 F_3^{(0)}$ and integrate it with the
resolvent. We have
\begin{eqnarray}&&
\hat{\cal L}_1 F_3^{(0)}=-C_3\Biggl\{
\frac{3(d-1)(3d-2)}{[r_{12}^2+r_{13}^2+r_{23}^2]^{d}}\nonumber\\&&-
\frac{d(d-1)}{[r_{12}^2+r_{13}^2+r_{23}^2]^{d+1}}\Biggl[
\frac{(r_{13}^2-r_{23}^2)^2}{r_{12}^2}-
\frac{(r_{12}^2-r_{23}^2)^2}{r_{13}^2}-
\frac{(r_{12}^2-r_{13}^2)^2}{r_{23}^2}\Biggr]
\nonumber\\&&+\frac{2d(d-1)^2}{[r_{12}^2+r_{13}^2+r_{23}^2]^{d+1}}\left(r_{12}^2\ln 
\frac{r_{13}^2 r_{23}^2}{r_{12}^4}+
r_{13}^2\ln \frac{r_{12}^2 r_{23}^2}{r_{13}^4}+
r_{23}^2\ln\frac{r_{12}^2 r_{13}^2}{r_{23}^4}\right)\Biggr\}
\nonumber\end{eqnarray}
There are three kinds of terms here. From the symmetry reasons, 
it is enough to know the values of the following integrals
\begin{eqnarray}&&
I_1 =\int \frac{d{\bbox x}_1d{\bbox x}_2d{\bbox x}_3}
{(x_{12}^2 +x_{13}^2 +x_{23}^2)^d \left[({\bbox x}_1-{\bbox y}_1)^2+
({\bbox x}_2-{\bbox y}_2)^2+({\bbox x}_3-{\bbox y}_3)^2\right]^{3d/2-1}} 
\nonumber\\&&
I_2 =\int \frac{(x_{13}^2 -x_{23}^2)^2 d{\bbox x}_1d{\bbox x}_2d{\bbox x}_3}
{x_{12}^2(x_{12}^2 +x_{13}^2 +x_{23}^2)^{d+1} 
\left[({\bbox x}_1-{\bbox y}_1)^2+
({\bbox x}_2-{\bbox y}_2)^2+({\bbox x}_3-{\bbox y}_3)^2\right]^{3d/2-1}}\nonumber
\\&&
I_3 =\int \frac{x_{12}^2 \ln \frac{x_{13}^2}{x_{12}^2}
d{\bbox x}_1d{\bbox x}_2d{\bbox x}_3}
{(x_{12}^2 +x_{13}^2 +x_{23}^2)^{d+1} \left[({\bbox x}_1-{\bbox y}_1)^2+
({\bbox x}_2-{\bbox y}_2)^2+({\bbox x}_3-{\bbox y}_3)^2\right]^{3d/2-1}}
\nonumber \end{eqnarray}
With the logarithmic accuracy one finds
\begin{eqnarray} &&
I_{1,2,3}=\frac{A_{1,2,3}}{\left(y_{12}^2+y_{13}^2+y_{23}^2\right)^{d-1}}
\ln\left[\frac{\sqrt{y_{12}^2+y_{13}^2+y_{23}^2}}{L}\right]\\&&
A_1=\frac{2\pi^{3d/2}}{3(d-1)\Gamma(3d/2-1)}\,,\quad
A_2=\frac{2\pi^{3d/2}}{3d(d-1)\Gamma(3d/2-1)}\,,\nonumber\\&&
A_3=-\frac{\pi^{3d/2}}{3d(d-1)\Gamma(3d/2-1)}\,.\nonumber
\end{eqnarray}
The result can be expressed in terms of $I_{1,2,3}$
\begin{eqnarray}&&
F_3^{(1)}=-\xi\hat{\cal L}_0^{-1}\hat{\cal L}_1 F_3^{(0)}=
-\xi\frac{C_3\Gamma(3d/2-1)}{2\pi^{3d/2}}
\left[(9d-6)I_1 -3dI_2 +2d(d-1)I_3\right]
\nonumber\\&&
=\frac{\xi C_3}{\left(y_{12}^2+y_{13}^2+y_{23}^2\right)^{d-1}}
\ln\left[\frac{L}{\sqrt{y_{12}^2+y_{13}^2+y_{23}^2}}{L}\right]=\xi F_3^{(0)}
\ln\left[\frac{L}{\sqrt{y_{12}^2+y_{13}^2+y_{23}^2}}\right]\nonumber
\end{eqnarray}
This is the first term of the expansion with respect to $\xi$ of the function
$F_3\propto (y_{12}^2+y_{13}^2+y_{23}^2)^{1-d-\xi}\propto r^{\gamma-1-d}$ which
is the normal scaling for the triple correlation function.

Analysis of the integrals $I_{1,2,3}$ shows that there is no angular
singularity at collinear geometry. By a direct calculation one can check that
the scaling of the triple correlation function is the same for three points on
a line. This is natural since nonzero $\gamma$ destroys degeneracy,
collinearity is no longer preserved during the Lagrangian evolution so the
correlation functions at collinear geometry have no anomaly, similarly to what
has been established by $\gamma$-expansion at small scales \cite{97BFL,97PSS}.

\section{Conclusion}

We have studied the correlation functions of a passive scalar in the
framework of the Kraichnan model on distances larger than the
scalar's pumping length. In the Batchelor limit, the collinear anomaly
has been found: scaling behavior of many-point correlation functions for the 
collinear geometry (where some points lie on a line) strongly differs from
one for general geometry. The anomalous scaling is observed in the
interval of angles which decreases with increasing scale. This violation
of a conventional scaling behavior is related to a strong correlation
between different Lagrangian trajectories occurring in the Batchelor case that is
for distances smaller than the viscous scale of the velocity field. For larger
distances (at the inertial interval of turbulence) the scale invariance of scalar
statistics (yet not Gaussianity) 
is likely to be restored (remember that we consider the scales larger
than the scale of scalar's pumping) as it is confirmed by our calculations in
Sect. \ref{ns}. At even larger scales (beyond velocity correlation scale that is
an external scale of turbulence) the scalar statistics has to be Gaussian.

\acknowledgements

We are grateful to Robert Kraichnan whose work is a permanent source of
inspiration for us.  We thank Alexander Zamolodchikov for stimulating
discussions. This work was supported by the Einstein Center at the
Weizmann Institute and by the grants of Minerva Foundation, Germany and Israel
Science Foundation.

\end{document}